# Toward a New Protocol to Evaluate Recommender Systems

Frank Meyer, Françoise Fessant, Fabrice Clérot
Orange Labs
av. Pierre Marzin
22307 Lannion cedex
France
{franck.meyer,francoise.fessant,fabrice.clerot}@orange.com

Eric Gaussier
University of Grenoble - LIG
UFR IM2AG - LIG/AMA
Grenoble Cedex 9
France
eric.gaussier@imag.fr

## ABSTRACT
In this paper, we propose an approach to analyze the performance and the added value of automatic recommender systems in an industrial context. We show that recommender systems are multifaceted and can be organized around 4 structuring functions: help users to decide, help users to compare, help users to discover, help users to explore. A global off line protocol is then proposed to evaluate recommender systems. This protocol is based on the definition of appropriate evaluation measures for each aforementioned function. The evaluation protocol is discussed from the perspective of the usefulness and trust of the recommendation. A new measure called Average Measure of Impact is introduced. This measure evaluates the impact of the personalized recommendation. We experiment with two classical methods, K-Nearest Neighbors (KNN) and Matrix Factorization (MF), using the well known dataset: Netflix. A segmentation of both users and items is proposed to finely analyze where the algorithms perform well or badly. We show that the performance is strongly dependent on the segments and that there is no clear correlation between the RMSE and the quality of the recommendation.

## Categories and Subject Descriptors
H.3.3 [**Information Search and Retrieval**]: Information filtering – *collaborative filtering, recommender system;* H.3.4 [**Systems and Software**]: Performance evaluation (efficiency and effectiveness) – *performance measures, usefulness of recommendation.*

## 1. INTRODUCTION
The aim of recommender systems is to help users to find items that should interest them, from large catalogs. One frequently adopted measure of the quality of a recommender system is accuracy (for the prediction of ratings of users on items) [1,14]. Yet in many implementations of recommender system services, the rating prediction function is either not provided, or not highlighted when it is provided (in industrial contexts, the generated recommendations themselves and their utility are more important than the rating predictions). There is increasing consensus in the community that accuracy alone is not enough to assess the practical effectiveness and added-value of recommendations [8,13]. Recommender systems in industrial context are multifaceted and we propose to consider them around the definition of 4 key recommendation functions which meet the needs of users facing a huge catalog of items: how to decide, how to compare, how to explore and how to discover. Once the main functions are defined, the next question is how to evaluate a recommender system on its various facets? We will review for each function the key points for their evaluation and the available measures if they exist. In particular, we will introduce a dedicated measure for the function "help to discover". This function raises the question of the evaluation from the point of view of the usefulness of the recommendation. We will also present a global evaluation protocol able to deal with the multifaceted aspect of recommender systems, which requires at least a simple segmentation of users and items. The remainder of the paper is organized as follow: the next section introduces the four core functions of an industrial recommender system. Then the appropriate measures for each core function are presented as well as the global evaluation protocol. The last part of the paper is dedicated to experimental results and conclusion.

## 2. MAIN FEATURES OF RECOMMENDER SYSTEMS
Automatic recommender systems are often used on e-commerce websites. These systems work in conjunction with a search engine for assistance in catalog browsing to help users find relevant content. As many users of e-commerce websites are anonymous, a very important feature is the contextual recommendation of item, for anonymous users. The purpose of these systems being also to increase usage (the audience of a site) or sales, the recommendation itself is more important than the rating predicted. Moreover, prioritizing a list of items on a display page is a more important functionality than the prediction of a rating. These observations, completed with interviews with marketers and project managers of Orange about their requirements relatively to recommender systems and an overview of recommender systems both in the academic and in the industrial fields [10] has led us to organize the recommender systems' functionalities into 4 main features:

**Help to Decide**. Given an item, a user wants to know if he will appreciate the item. This feature consists of the prediction of a rating for a user and an item and is today mainstream in academic literature [14].

**Help to Compare**. Given several items, a user wants to know what item to chose. This feature corresponds to a ranking function. It can be used to provide recommendation lists [5] or to provide personalized sorting results of requests on a catalog.

**Help to Discover**. Given a huge catalog of items, a user wants to find a short list of new interesting items. This feature is usually called item-based top-N recommendation in the academic literature [6]. It corresponds to personalized recommendation. Note that the prediction of the highest rated item is not necessarily the most useful recommendation [5]. For instance the item with the highest predicted rating will most likely be already known by the user.

**Help to Explore (or Navigate)**. Given one item, an (anonymous) user wants to know what the related items are. This feature corresponds to the classical item-to-item recommendation to anonymous users popularized by the e-commerce website Amazon [9] during catalog browsing. This function is widely used in the industry because it can make recommendations for anonymous users, based on the items she consults. It requires a similarity function between items.

# 3. EVALUATION OF INDUSTRIAL RECOMMENDER SYSTEMS

In this section we discuss the appropriate measures for each core function and a global protocol for the evaluation of the recommender system. The evaluation is viewed from the standpoint of the utility of the recommendation for each user and each item.

## 3.1 Utility of the recommendation

A good recommender system should avoid bad and trivial recommendations. The fact that a user likes an item and the fact that an item is already known by the user have to be distinguished [7]. A good recommendation corresponds to an item that would probably be well rated by the user but also an item that the user does not know. For instance it is worthless recommending to all users the blockbuster of the year: it should be a good rated movie on the average, but it is not a useful recommendation as most of people may have already seen it.

## 3.2 Item segmentation and user segmentation

Another important issue for an industrial application is to fully exploit the available catalog, including its long tail, consisting of items rarely purchased [2]. A system's ability to make a recommendation, in a relevant way, for all items in the catalog is therefore important. However Tan and Netessine [16] have observed on the Netflix dataset for instance, that the long tail effect is not so obvious. There's more of a Pareto distribution (20% of the most rated items represents 80% of the global ratings) in the Netflix data than a long tail distribution as proposed by Anderson [2] (where infrequent items globally represent more ratings). They also noticed that the behavior of the users and the type of items they purchase are linked. In particular, customers who watched items in the long tail are in fact heavy users, light users tend to focus only on popular items. These observations lead us to the introduction of the notion of segments of items and users. The definition of the segment thresholds must be relative and catalog dependant. We will use the terms of light/heavy users segment and of unpopular/popular item segment instead of using long tail and short head concepts. In a first step we will use this simple segmentation to analyze how an industrial recommender system can help all users both heavy and light and how it can recommend all items, both popular and unpopular.

## 3.3 Measures of performance

For our protocol we use a classic train/test split of the data. The train set will be used to compute statistics and thresholds and to build a predictive model. The test set will be used to compute the performance measures. The predictive model should at least be able to provide a rating prediction function for any couple of user and item. We will see that to provide the "Help to Explore" functionality the predictive model also must be able, in some way, to produce an item-item similarity matrix allowing it to select, for each item $i$, its most similar items (the related items). We first detail the performance measures we use for our protocol, according to the 4 core functions.

**Help to Decide**. The main use case is a user watching an item description on a screen and wondering if he would enjoy it. Giving a good personalized rating prediction will help the user to choose. The "help to decide" function can be given by the rating prediction function and must be measured by an accuracy measure which penalizes extreme errors. The Root Mean Squared Error (RMSE) is the natural candidate [14].

**Help to Compare**. The main use case here is a user getting an intermediate short list of items after having given her preferences. This user then wants to compare the items of this short list, in order to choose the one she will enjoy most. The function needs a ranking mechanism with a homogeneous quality of ranking over the catalog. A simple measure is the percentage of compatible rank indexes. After modeling, for each user $u$ and for each couple of item $(i, j)$ in the test set rated by $u$ with $r_{u,i} \neq r_{u,j}$, the preference given by $u$ is compared with the predicted preference given by the recommender method, using the predicted ratings $\hat{r}_{u,i}$ and $\hat{r}_{u,j}$. The percentage of compatible preferences is given by:

$$comp = \frac{\sum_{\{(r_{u,i}) \in T, (r_{u,j}) \in T, r_{u,j} \neq r_{u,i}\}} c(r_{u,i}, r_{u,j})}{|\{(r_{u,i}) \in T, (r_{u,j}) \in T, r_{u,j} \neq r_{u,i}\}|} \quad (3\text{-}1)$$

with $c(r_{u,i}, r_{u,j}) = \delta\left(sign(r_{u,i} - r_{u,j}), sign(\hat{r}_{u,i} - \hat{r}_{u,j})\right)$, where $\delta\left(sign(r_{u,i} - r_{u,j}), sign(\hat{r}_{u,i} - \hat{r}_{u,j})\right)$ is 1 if $r_{u,i} - r_{u,j}$ has the same sign as $\hat{r}_{u,i} - \hat{r}_{u,j}$ and 0 otherwise, and $|\{(r_{u,i}) \in T, (r_{u,j}) \in T, r_{u,j} \neq r_{u,i}\}|$ is the number of elements of $\{(r_{u,i}) \in T, (r_{u,j}) \in T, r_{u,j} \neq r_{u,i}\}$.

**Help to Discover**. The main use case here is a user getting recommended items: these recommendations must be relevant and useful. For relevancy our approach is the following: an item $i$ recommended for the user $u$

- is considered **relevant** if $u$ has rated $i$ in the test set with a rating greater than or equal to u's mean of ratings,

- is considered **irrelevant** if $u$ has rated $i$ in the test set with a rating lower than u's mean of ratings

- is not evaluated if not present for $u$ (not rated by $u$) in the test set.

The classical measure to evaluate recommendation list is the precision measure (recall being difficult to apply in the context of recommendation, as in huge catalogs one does not know all the items relevant for each user). For each user u:

$$\text{precision}_u = \frac{\text{number of relevant recommended items}_u}{|H_u|} \quad (3\text{-}2)$$

$H_u$ stands for the subset of evaluable recommendations in the test set for u, that is to say the set of couples $(u,i)$, $i$ being the recommended item to the user $u$. $|H_u|$ is the size of $H_u$, in number of couples $(u, i)$.

However the precision is not able to measure the usefulness of the recommendations: recommending well-known blockbusters, already known by the user will lead to a very high precision although this is of very low utility. To account for this, we introduce here the concept of recommendation impact. The basic idea is that, the more frequent a recommended item is, the less impact the recommendation has. This is summarized in Table 1:

Table 1. The notion of recommendation Impact

|  | Impact of the recommendation | |
|---|---|---|
|  | **Impact** if the user **likes** the item | **Impact** if the user **dislikes** the item |
| Recommending a popular item | **Low**: the item is likely to be already known at least by name by the user. | **Low**: even if the user dislikes this item he can understand that as a popular item this recommendation is likely to appear... at least at the beginning |
| Recommending an unpopular (infrequent) item | **High:** the service provided by the recommender system is efficient. The rarest the item was, the less likely the user would have found it alone. | **High:** not only the item was unknown and did not inspire confidence, but it also was not good. |

We then define the Average Measure of Impact (AMI) for the performance evaluation of the function "Help user to Discover". The AMI of a recommendation list $Z$ for a user $u$ with an average of rating $\bar{r}_u$ is given by:

$$AMI_u(Z) = \frac{1}{|H_u|}\sum_{(u,i)\in Z, (u,i)\in H} \frac{1}{count(i)} \times sign(r_{u,i} - \bar{r}_u) \times |I| \quad (3\text{-}3)$$

Where $H_u$ denotes the subset of the evaluable recommendations in the test set, $Z$ denotes the set of couples (user, item), representing a set of recommendations, count(i) the number of logs in the train set related to the item i, and $|I|$ the size of the catalog of items.

The rarer an item $i$ (rarity being estimated in the train set), the greater the AMI if $i$ is both recommended and relevant for a user $u$. The greater the AMI, the better the positive impact of the recommendations on $u$. The AMI will have to be calibrated as we do not know yet what is a "good AMI". But we can already compare different algorithms, or different recommendation strategies (such as post filtering methods to add serendipity) with this measure.

**Help to Explore.** The main case here is the item-to-item recommendation for an anonymous user who is watching an item description on a screen: the recommender system should propose items similar to that being watched. We can try to evaluate the performance of this functionality by associating, with each context item $i$, the KNN of $i$, using an overall precision measure for the recommended items. But, we will have an issue: it can be more effective to associate each context item $i$ with $N$ items optimized only for precision, rather than $N$ items similar to the context item $i$. It may be more efficient, to optimize precision, to associate blockbusters for each source item. In fact we want to assess the quality of the Help to Explore (navigate) function: we want a good semantic, meaningful similarity for each associated item. But only an experiment with real users can assess this semantic similarity.

Our solution is to use the underlying item-item similarity matrix for this evaluation. We can assess the overall quality of the pairs of similar items by an indirect method: 1. given a predictive model, find a way to compute similarities between any pair of items, building an item-item similarity matrix. 2. use an item-item K-Nearest Neighbors (KNN) model [12] using this matrix. The assumption is that a good similarity matrix must lead to good performances for other aspects of the recommendation when used into an item-item KNN model. This is the approach we take, using RMSE, precision, and ranking performance measures. For a KNN type algorithm, this analysis is straightforward and simple: the similarity matrix is already the kernel of the model. The algorithms that are not directly based on a similarity measure need a method for extracting the similarities between the items. For matrix-factorization-based algorithm, this can correspond to a method to compute similarities between the factors of the items.

### 3.4 Evaluation Protocol

The evaluation protocol is then designed thanks to the mapping between the 4 core functions and the associated performance measures as summarized in Table 2.

Table 2. Adapted measures for each core function

| Functions | Quality criterion | Measure |
|---|---|---|
| Decide | Accuracy of the rating prediction | RMSE |
|  | Penalization of extreme errors to minimize the risk of wrong decision |  |
| Compare | Good predicted ranking for every couple of items of the catalog | COMP<br>% of compatible rank indexes |
| Discover | Selection for a user the most preferred items in a list of items | (Precision, not recommended!) |
|  | Identification of good/bad recommendations | Average Measure of Impact (AMI) |
|  | Precise, useful, trusted recommendation |  |
| Explore | Precise recommendations | Similarity matrix leading to good performances, in accuracy, relevancy, usefulness and trust |
|  | Identification of good/bad recommendations |  |

The following notations are adopted: a log *(u, i, r)* corresponds to a user *u* who rated an item *i* with the rating *r*. *U* is the set of all the users, *I* is the set of all the items. Given a dataset *D* of logs and an algorithm *A*, the evaluation protocol we propose is as follow:

**Initialization**

Randomly split the dataset into 2 datasets train and test

Use the train dataset to generate a model with the algorithm A.

**Evaluation**

1. For each log (u, i, r) of the test set:

    1.1 compute the predicted rating of the model

    1.2 compute the predicted rating error

2. Use the RMSE which gives an indicator of the performance of the Help to Decide function.

3. For each user u of U:

    3.1 sort all u's logs of the test set by ratings

    3.2 sort all u's logs of the test set by rating prediction

    3.3 compute COMP comparing the indexes of u's logs and the indexes of the predicted ratings of he logs.

4. Use the averaged COMP as an indicator of the Help to Compare function.

5. For each item i of I, compute count(i) which is the number of logs in the train set referencing i.

6. For each user u of U:

    6.1 compute the predicted rating of each item i of I.

    6.2 select the top-N highest predicted rating items noted $i_{u,1}$ to $i_{u,N}$ which are the Top-N recommended items.

    6.3 compute the rating average of u, noted $\bar{u}$.

    6.4 for each recommended item $i_{u,j}$ of u:

        6.4.1 check if a corresponding log (u, $i_{u,j}$,r) exists, If so the recommendation of $i_{u,j}$ is evaluable else skip the step 6.4.2.

        6.4.2. If $r \geq \bar{u}$ then the recommendation is considered relevant (and irrelevant in the other case).

    6.5 compute the Precision and the AMI for the evaluable recommendations

7. Use the Precision and the AMI, averaged by users, as the indicators for the Help to Discover Function

8. Specify a way to compute efficiently, using the model of the algorithm A, the similarity between every couple of items (i,j).

9. Compute the similarity matrix of all the couple (i, j) for I×I.

10. Use this similarity matrix as the kernel of an item-item K-Nearest Neighbor model, then run the protocol for the steps 1 to 7 for RMSE, COMP, AMI and Precision to obtain a 4-dimensional indicator of the quality of the Help to Explore function.

## 4. EXPERIMENTS
## 4.1 Datasets and configuration

Experiments are conducted on the widely used dataset Netflix [3]. This dataset has the advantage of being public and allows performance comparisons with many others techniques. Agnostic thresholds are used for segments of users and items, depending of datasets. We used simple thresholds based on the mean of the number of ratings to split items into popular items and unpopular (infrequent) items, and similarly to split users into heavy users and light users. For instance, on Netflix, using a Train Set of 90% of the total of logs, the mean of the number of rating for the users is 190 (heavy users are users who gave more than 190 ratings otherwise they are light users) and the mean of number of ratings for the items is 5089 (popular items are items with more than 5089 ratings otherwise they are unpopular items). The number of generated items for the Top-N recommendation is always N=10. All our tests are carried out on this configuration: Personal Computer with 12 GB Ram, processor Intel$^{TM}$ Xeon$^{TM}$ W3530 64-bit-4-core processor running at 2.8 GHz, hard disk of 350 GB. All algorithms and the benchmark process are written in Java$^{TM}$.

## 4.2 Algorithms

We chose to use 2 models: fast matrix factorization using the MF algorithm presented in [15] and an item-item KNN algorithm [12]. These algorithms are mainstream techniques for recommender systems. For MF we analyze the effect of the number of factors, for the KNN algorithm we analyze the effect of K, the number of Nearest Neighbor kept in the model. In addition, to compare the performances of these 2 algorithms, 2 baseline algorithms are also used:

- a simple default predictor using the mean of items and the mean of the users (the sum of the two means if available, divided by 2). This algorithm is also used by the KNN algorithm when no KNN items are available for a given item to score.

- a random predictor, generating uniform ratings between [1..5] for each rating prediction.

One industrial requirement of our system was that it could take into account new items and new users every 2 hours. Considering other process and I/O constraints, for all the algorithms the modeling time was then restricted to 1.5 hours. This has implications for the MF algorithm as on Netflix it always reaches an optimum between 16 and 32 factors: this is a constant for all our tests, for all the performances. Beyond 32 factors, MF does not have enough time to converge. Note that this convergence may be slow, longer than 24 hours for more than 100 factors on the Netflix dataset.

**Implementations details**

Our implementation of MF is similar to those of the BRISMF implementation [15] with a learning rate of 0.030 and a regularization factor of 0.008, with early stopping. Learning process is stopped after 1.5 hours, or when the RMSE increases three consecutive times (the increase or decrease of the RMSE is controlled on a validation set consisting of 1.5% of the train set). We used an implementation of item-item KNN model as described in [11]. The similarity function is the Weighted Pearson similarity [4]. All details about implementations can be found in [10].

## 5. NUMERICAL RESULTS

The following abbreviations are used for the segmentation of the performance: Huser: Heavy users, Luser: Light users, Pitem: Popular items and Uitem: Unpopular items (the meaning of unpopular is rather "rare", "infrequent"). For MF we analyzed the number of factors used and for KNN the number of NN kept. The full results of our experiments are available in [10].

## 5.1 "Help to Decide" performances

The global default predictor has a RMSE of 0.964 and the global random predictor has a RMSE of 1.707.

**KNN's RMSE performances:** Different sizes of neighborhoods (K) have been tested, compliant with our tasks in an industrial context. Increasing K generally increases the performances. However the associated similarity matrix weights must be kept in RAM for efficiency purposes, which is difficult, if not possible, with high values of K. For very large catalog applications, the size of the KNN matrix must be reasonable (up to 200 neighbors in our tests). The KNN method performs well except when K is small and except for the light-user-unpopular item segment (Luser Uitem). There is a significant gap between the RMSE for the LuserUitem segment (RMSE=1.05) and the RMSE of the heavy-user-popular-item segment (RMSE=0.8). Clearly, the KNN model is not adapted to the former, whereas it performs well on the later. Optimal number of neighbors is around K= 100.

**MF's RMSE performances:** Different numbers of factors have been tested. MF has difficulties modeling the Luser-Uitem segment: on this segment the RMSE never decreases under 0.96. On the contrary the RMSE for heavy-user-popular-item is close to 0.81, and the two symmetrical segments light-user-popular item and heavy-user-unpopular-item both have a good (low) RMSE (0.84 and 0.85). The RMSE decreases when number of the factor increases up to around 20 factors. After that number, the RMSE increases. It is a consequence of our time-constrained early stopping condition. This corresponds to about 140 passes on the train set. The optimal number of factors seems to be between 16 and 32.

## 5.2 "Help to Compare" performance

The default global predictor has a percentage of compatible rank indexes (COMP) of 69% and the random global predictor has a performance of 49.99%.

**MF's and KNN's ranking performances:** The results are given for the time limited version of run for MF. MF outperforms the KNN model for the light user segments (with a COMP of 73.5% for MF and 66% for KNN). For the rest, the performances are similar to those of KNN. The maximum of ranking compatibility is around 77% for heavy users' segments.

## 5.3 "Help to Discover" performance

### 5.3.1 Analysis using the Precision

The global default predictor has a precision of 92.86 % which is questionable: one can see that a simple Top-10 based on high rating average is sufficient to obtain good precision performance. The global random predictor has a precision of 53.04%.

**KNNs' precision performances:** The precision increases as the number of K increases. But the results are not significantly better than that of the default predictor. The precision is better than the default predictor for only the Huser-Pitem segment and only for at least K=200. Under K=100, it seems better to use a default predictor than a KNN predictor for ranking tasks. Nevertheless the Huser-Pitem segment is well modeled: the precision for 10 generated items for the KNN model is greater than 97% for the model with 200 neighborhoods.

**MF's precision performances:** MF has a better behavior than the KNN model, especially for the light-user-unpopular-item segment (precision of 96% for F=32 factors, precision of 83% for the KNN with K>=100).

### 5.3.2 Analysis using the AMI

The Average Measure of Impact gives slight negative performances for the random predictor and a small performance to the default predictor: the default predictor "wins" its impact values on Unpopular items. Note that the supports for the different evaluated segments are very different and the weights of the two popular item segments are significantly higher The KNN model behaves significantly better that the default predictor for the AMI. For MF, the behavior is much worse than that a KNN model. In general, the impact of MF is similar to, or lower than that of the default predictor. An analysis according to the segmentation gives a more detailed view of where are the impacts. Numerical results are summarized in Table 3.

**Table 3. AMI according to the segmentation**

| Best model | Huser Pitem | Luser Pitem | Huser Uitem | Luser Uitem | Global |
|---|---|---|---|---|---|
| MF F=32 | 0.38 | 0.26 | 8.93 | 10.61 | **0.5** |
| KNN K=100 | **0.71** | **0.43** | 9.59 | 8.84 | **2.0** |
| Default Pred | 0.29 | 0.25 | **21.22** | **12.31** | **0.5** |
| Random Pred | 0.00 | 0.03 | -5.13 | -0.53 | **-0.6** |
| Best algorithm | KNN | KNN | Default Predictor | Default Predictor | **KNN** |

## 5.4 Summary for Decide, Compare, Discover

Four models have been analyzed: a KNN model, a MF model, a random model and a default predictor model, on 3 tasks adapted to a rating-predictor-based recommender system: Decide, Compare, Discover and on 4 user-item segments: heavy-user-popular-item, heavy-user-unpopular-item, light-user-popular-item and light-user-unpopular item. A summary of the results is given in Table 4. An analysis of the results by segments shows that globally, KNN is well adapted for the heavy-user segments and that MF, and the default predictor are well adapted to light-user segments. Globally, for the tasks "Help to Decide" and "Help to Compare", MF is the best-suited algorithm in our tests. For the task "Help to Discover" KNN is more appropriate. Note that a switch-based hybrid recommender [14], based on item and user segmentation could exploit this information to improve the global performances of the system. Finally 3 main facts will have to be considered:

1. Performances strongly vary according to the different segments of users and items.

2. MF, KNN and default methods are complementary as they perform differently across the different segments.

3. RMSE is not strictly linked to other performance measure, as mentioned for instance in [5].

**Table 4. Global results, summary**

|  | Heavy Users Popular items | Heavy Users Unpopular items | Light Users Popular Items | Light Users Unpopular Items |
|---|---|---|---|---|
| Decide RMSE | KNN | MF | MF | MF |
| Compare %Compatible preferences | KNN | KNN | MF | MF |
| Discover Precision | KNN | MF | Default Predictor | MF |
| Discover Average Measure of Impact | KNN | Default Predictor | KNN | Default Predictor |

When designing a recommender engine, we have to think about the impact of the recommender: recommending popular items to heavy users might be not so useful. On the other hand, it can be illusory to make personalized recommendations of unpopular (and unknown) items to light (and unknown) users. A possible simple strategy could be:

- rely on robust default predictors, for instance based on robust means of items to try to push unknown golden nuggets to unknown users,
- use personalized algorithms to recommend popular items to light users,
- finally, use personalized algorithms to recommend unpopular items of the long tail for heavy "connoisseurs".

## 5.5 "Help to Explore" performance

To analyze the performance of the "Help to Explore" functionality we have to compare the quality of the similarities extracted from the models. We use the protocol defined before: a good similarity matrix for the task "Help to Explore" is a similarity matrix leading to global good performances, when used in a KNN model. We choose a similarity matrix with 100 neighbors for each item: this is largely enough for item-to-item tasks where generally a page displays 10 to 20 similar items. Results are presented in Table 5 for the KNN models with K=100, comparing KNN computed on

MF's items factors, native KNN and a Random KNN model used as baseline. As item-item similarity matrix is the kernel of a item-item KNN model, compute similarities in this case is straightforward. To compute similarities between items for MF, we use the MF-based representation of items (the vectors of the factor of the items), with a Pearson similarity. The KNN model computed on the MF's factors of the items can be viewed as a MF-emulated KNN model. Note that as the default predictor model based on items' means and users' means cannot itself produce a similarity matrix, it is disqualified for this task. For the RMSE, the MF-Emulated KNN model looses 0.025 point going from 0.844 to 0.870. Compared with other models, it still performs correctly.

**Table 5. Quality of an item-item similarity matrix according to 4 measures: results on Netflix**

|  | Native KNN | KNN computed on MF's items factors |
|---|---|---|
|  | K=100 | K=100, number of factors=16 |
| RMSE | 0.8440 | 0.8691 |
| Ranking: % compatible | 77.03% | 75.67% |
| Precision | 91.90% | 86.39% |
| AMI | 2.043 | 2.025 |
| (Global time of the modeling task) | (5290 seconds) | (3758 seconds) |

For the global ranking, the difference between the MF-Emulated model and the native KNN model is still low, whereas a random KNN model performs very badly. For the precision, for a Top-10 ranking, the MF-Emulated KNN model performs significantly worse than a native KNN model. For the Average Measure of Impact, the MF-emulated KNN model and the native KNN model perform almost identically. These results show that MF could be used to implement a similarity function between items to support the "Help to Explore" function, and that MF could be used as a component for faster KNN search.

## 6. CONCLUSION

We have proposed a new approach to analyze the performance and the added value of automatic Recommender Systems in an industrial context. First, we have defined 4 core functions for these systems, which are: Help users to Decide, Help users to Compare, Help users to Discover, Help users to Explore. Then we proposed a general off-line protocol crossing our 4 core functions with a simple 4 users×items segments to evaluate a recommender system according to the industrial and marketing requirements. We compared two major state of the art methods, item-item KNN and MF, with 2 baselines methods used as reference. We showed that the two major methods are complementary as they perform differently across the different segments. We proposed a new measure, the Average Measure of Impact, to deal with the usefulness and the trust of the recommendations. Using the precision measure, and the AMI, we showed that there is no clear evidence of correlation between the RMSE and the quality of the recommendation. We have demonstrated the utility of our protocol as it may change

- the classical vision of the recommendation evaluation, often focused on the RMSE/MAE measures as they are assumed correlated with the system overall performances,

- and the way to improve the recommender systems to achieve their tasks.